\documentclass[a4paper,11pt]{article}
\usepackage{pos}
\usepackage{enumitem}

\title{Monte Carlo Simulations and Validation of NectarCAM, a Medium Sized Telescope Camera for CTA}
 \ShortTitle{Monte Carlo Simulations of NectarCAM}

\author*[a]{Thomas P. Armstrong}
\author[a]{Heide Costantini}
\author[b]{Jean-François Glicenstein}
\author[c]{Jean-Philippe Lenain}
\author[d]{Ullrich Schwanke}
\author[b]{Thomas Tavernier}

\affiliation[a]{Aix Marseille Univ, CNRS/IN2P3, CPPM, Marseille, France}

\affiliation[b]{CEA-IRFU, Gif-Sur-Yvette, France}

\affiliation[c]{LPNHE, CNRS/IN2P3, Paris, France}

\affiliation[d]{Humboldt University, Department of Physics, Berlin, Germany}

\forColl{CTA} 

\emailAdd{armstrong@cppm.in2p3.fr}

\abstract{The upcoming Cherenkov Telescope Array (CTA) ground-based gamma-ray observatory will open up our view of the very high energy Universe, offering an improvement in sensitivity of 5-10 times that of previous experiments. NectarCAM is one of the proposed cameras for the Medium-Sized Telescopes (MST) which have been designed to cover the core energy range of CTA, from 100 GeV to 10 TeV. The final camera will be capable of GHz sampling and provide a field of view of 8 degrees with its 265 modules of 7 photomultiplier each (for a total of 1855 pixels). In order to validate the performance of NectarCAM, a partially-equipped prototype has been constructed consisting of only the inner 61-modules. It has so far undergone testing at the integration test-bench facility in CEA Paris-Saclay (France) and on a prototype of the MST structure in Adlershof (Germany). To characterize the performance of the prototype, Monte Carlo simulations were conducted using a detailed model of the 61 module camera in the CORSIKA/sim\_telarray framework. This contribution provides an overview of this work including the comparison of trigger and readout performance on test-bench data and trigger and image parameterization performance during on-sky measurements.}

\FullConference{37$^{\rm{th}}$ International Cosmic Ray Conference (ICRC 2021)\\
		July 12th -- 23rd, 2021\\
		Online -- Berlin, Germany}

\begin{document}
\maketitle

\section{Introduction}

CTA is the next generation of ground based imaging atmospheric telescopes, representing the move from current experiments to a full observatory. With its two arrays covering both hemispheres, it will provide an order of 5-10 times improvement in sensitivity. Using three different sizes of telescope, it will be able to observe gamma rays with energies between 20~GeV to over 300~TeV. The bulk of this energy range will be met with the 12~m diameter Medium Sized Telescope (MST). NectarCAM is one of the proposed cameras which will be mounted on the MST \cite{NectarCAM}. Its concept is based on a modular design, where a module consists of 7 photomultiplier tubes (PMTs) and an associated set of readout and trigger electronics. The full camera will consist of 1855 PMTs providing an 8~degree field of view. The readout is based on the NECTAr ASIC which is able to store data in a circular buffer with GHz sampling until the camera is triggered (resulting in a 60~ns readout window). The camera has two gain channels with the nominal voltage able to measure the single photo-electron (p.e.) level and a higher gain providing a dynamic range up to 2000 p.e. (with a linearity of 5\%). 

The trigger logic of NectarCAM uses a multi-level scheme in order to reduce the number of random triggers from noise, while maximising the number of shower images recorded. In the first step, or the Level 0 (L0) trigger, a copy of the analog signal from an individual pixel is sent to the L0 ASIC, where the signal is compared to a programmable voltage threshold using a discriminator circuit. The output of the discriminator consists of gate pulse, reshaped to a programmable gate width at the trigger FPGA which also handles the Level 1 (L1) trigger fabric. The L1 trigger is based on the processing of L0 signals of overlapping 37-pixels regions, where the signal from each 7-pixel module is shared with its 6 neighbours. Several trigger algorithms can be implemented, but the default 3 Nearest Neighbours (3NN) is currently used in NectarCAM. 

A first demonstrator prototype of NectarCAM has been constructed and evaluated. Consisting of only the central 427 PMTs, this prototype has facilitated full testing and validation of the camera concept. The majority of the tests have been carried out at the CEA Paris-Saclay dark room test bench (France), a 12~m long dark room which is equiped with a LED pulser and continuous Night Sky Background (NSB) emulating light. To allow further tests, including integration with the telescope structure, the demonstrator was mounted on the prototype MST located in Adlershof (Berlin, Germany) where on-sky observations were carried out. 

The performance of CTA is estimated using Monte Carlo (MC) simulations of particle air showers produced by gamma rays and background protons and electrons. The response to the resulting Cherenkov light through the telescope optics and camera electronics therefore needs to be well understood. A large effort has gone into ensuring the models for each telescope are accurate, through performing matching simulations to tests carried out in the lab and on-sky. In this paper, a summary of the results from this process with NectarCAM will be presented, covering dark room tests in Section \ref{sec:testbench} and on-sky tests in Sections \ref{sec:adlershof}.

\section{Simulation of Test Bench}
\label{sec:testbench}

For the work presented in this paper, the simulation software \texttt{CORSIKA} (v6.9) and \texttt{sim\_telarray} (2018-11-07) are used \cite{corsika,simtel}. A light source similar to the LED flasher is implemented in the simulations including the wavelength (405~nm), the light pulse shape (Gaussian with a standard deviation of 0.64~ns) and the angular distribution (flat response over 11.4 degrees opening angle). The model for the camera used in the \texttt{sim\_telarray} simulations was updated with various lab measurements and design specifications. Some model parameters were adjusted during the comparisons presented in this section and are mainly parameters which are difficult to measure directly in a lab. While there are many intermediate results in the validation process, the two main outcomes presented here are the validation of the readout in terms of the intensity resolution, and the matching of the trigger performance for both the L0 and L1 stages.

\subsection{Verification of Readout Performance}

\begin{figure}[t]
\begin{centering}
\includegraphics[width=0.6\textwidth]{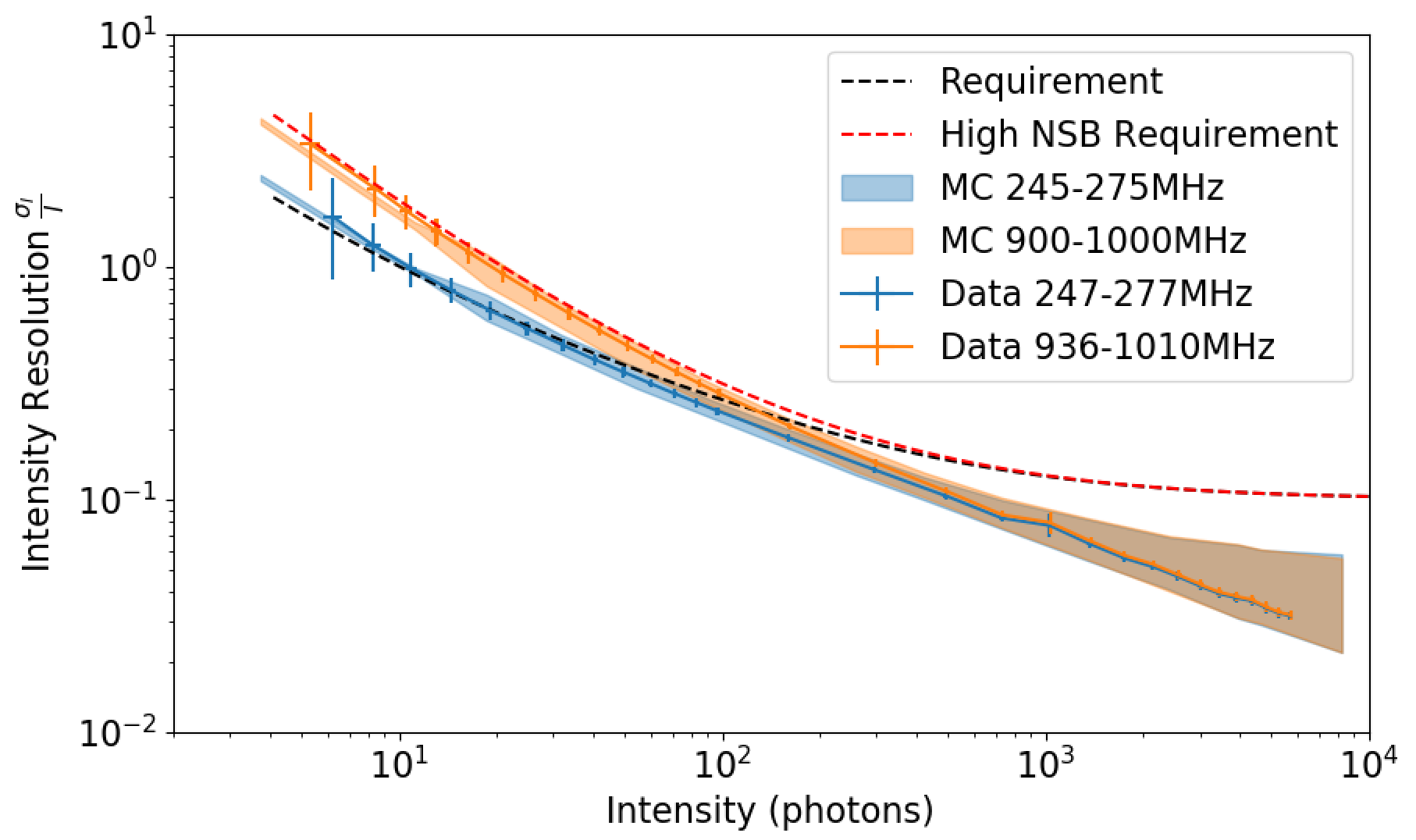}
\caption{The derived intensity resolution for nominal NSB ($\sim$250~MHz) in blue and high NSB ($\sim$1GHz) in orange, where the ranges are due to the uncertainty of the NSB test bench source. Also shown is the CTA requirements on the intensity resolution for these two levels.}
\label{fig:chargeres}
\end{centering}
\end{figure} 

The reconstructed charge was studied at the CEA dark chamber with a LED flasher and the NSB source. Before the measurement was made, the linearity of the response was tested and the cross-talk between pixels was measured to be negligible. Data were taken at a range of values of NSB between 0 - 1~GHz. The data were calibrated using the pedestals obtained without light sources (dark events) and the measured gain derived from the single p.e. spectrum, obtained using the method described in Ref. \cite{icrc2019}. For the simulations, a data set of a 1000 events was created for each illumination in the range of sub p.e. to greater than 2000~p.e.. The camera simulation was performed with a range of NSB values chosen in the same range as the data. The gain and pedestal values are generated automatically by \texttt{sim\_telarray}.


Both the data and the simulations were processed using the prototype processing pipeline for CTA, \texttt{ctapipe} (version v0.8.0) \cite{ctapipe},  where the charge was extracted using an integration window of width of 16~ns starting 6~ns before the peak of the signal. This was chosen to encompass the full pulse width recorded in the waveform. In order to include the effect of the photon detection efficiency (PDE) in the results, the values in p.e. were converted back to photons, using a conversion factor of 3.73~ph~p.e.$^{-1}$ derived from the total camera efficiency at 405~nm (the wavelength of the flasher). The intensity resolution was then calculated, using the expression

\begin{equation}
\frac{\sigma_I}{I} = \frac{\sqrt{\frac{1}{N} \Sigma_{t=1} (I_{rec,i} - \bar{I}_{rec})^2 }}{\bar{I}_{rec}},
\end{equation}
where $I_{rec,i}$ is the intensity measured for the $i$th event and $\bar{I}_{rec}$ is the mean over events. This form of intensity resolution does not take into account any bias present in the reconstruction. This requires an absolute calibrated light source which was not available. The results for two separate NSB regimes are shown in Figure \ref{fig:chargeres}, where these represent the nominal dark sky conditions and high NSB conditions. From here it can be seen that the simulations and data match well, providing confidence in the model constructed for the camera.

\subsection{Verification of Trigger Performance}
\label{sec:trig}

\begin{figure}[t]
\begin{centering}
\begin{tabular}{cc}
\includegraphics[width=0.44\textwidth]{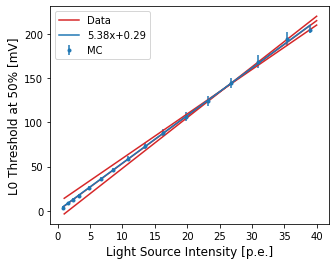} &
\includegraphics[width=0.45\textwidth]{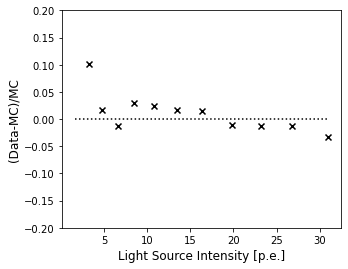} \\
\end{tabular}
\caption{\textbf{Left:} L0 threshold as a function of illumination, where the black lines represent the dispersion of measurements from different pixels taken in the lab and the blue is the simulations. \textbf{Right:} The difference between the L1 50\% trigger threshold values for data and simulations for the same illumination. }
\label{fig:trig}
\end{centering}
\end{figure}

The trigger efficiency was studied with the same dark chamber set up as with the readout study but excluding the use of the NSB source. The gains of each pixel were ajusted based on flat fielding measurements. For each illumination, data were recorded with a range of trigger threshold levels ensuring that the transition from 100\% to 0\% trigger rate was recorded. For the simulation, a dataset of 1000 events was created for each illumination level and trigger threshold from sub p.e. to $\sim$40~p.e., the same used in the test bench, and the trigger efficiency was recorded.

In a first step, only the performance of the L0 trigger was considered, i.e. the trigger efficiency of each pixel separately. In each scan, the trigger threshold which provided a 50\% trigger efficiency was recorded. The results are shown in the left panel of Figure \ref{fig:trig} where it can be seen that the relationship between expected light level in p.e. and the L0 threshold in digital counts at 50\% trigger efficiency matches the data very well. 
%

In a second step, simulations were performed using a single module of 7 pixels with a 3NN (next neighbours) trigger condition in order to evaluate the L1 performance. The right panel of Figure \ref{fig:trig} shows the comparison of the L1 threshold at 50\% trigger efficiency for data and MC with matching illumination levels. It can be seen that the trigger level matches well (within 5\%) further providing confidence in the simulation model.

\section{Simulation of Test Observations at Adlershof}
\label{sec:adlershof}

\begin{figure}
\centering
\includegraphics[width=\textwidth]{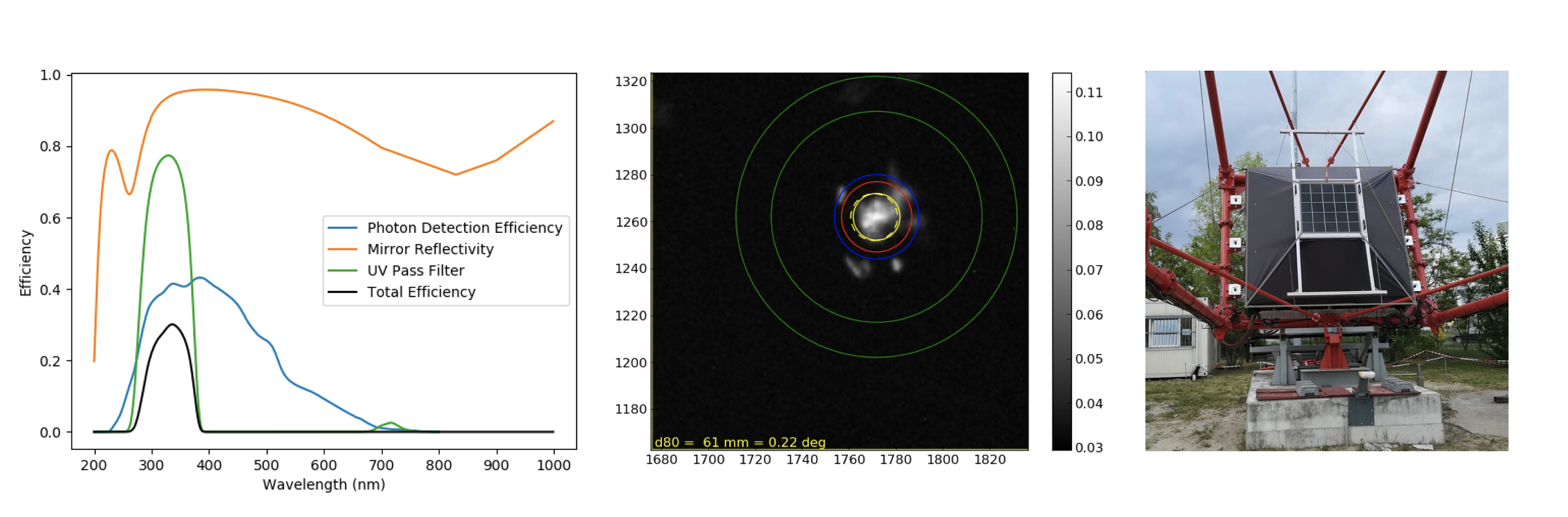} 
\caption{\textbf{Left:} Wavelength dependent efficiency model parameters, including the quantum efficiency, the mirror reflectivity, the UV Pass filter along with the total convolved efficiency. \textbf{Middle:} Example image of star from which the on-sky PSF was extracted. A non-smooth distribution of light is present due to misaligned mirror segments. The blue circle is the 80\% containment radius, the green lines denote the ring used for background subtraction; the yellow (solid + dashed) circles denote two different ways to estimate the centre of gravity.  \textbf{Right:} An image of the baffle mounted in front of the camera in order to reduce the NSB.}
\label{fig:fig3}
\end{figure}

In May and June of 2019, tests were performed with the NectarCAM demonstrator mounted on the MST prototype structure in Adlershof. While the main goal was to perform integration tests, several changes were implemented to reduce the NSB contribution from the surrounding light pollution, resulting in the successful observation of air-shower events. To test the model constructed for the camera in realistic conditions, simulations were performed to compare to the air-shower data observed. However, before this could be done, several changes had to be made. 

\begin{itemize}[noitemsep]
\item UV Filter - To help reduce the level of NSB, a UV Pass filter was placed in front of the camera. The transmission as a function of wavelength can be seen in the left panel of Figure \ref{fig:fig3} along with the PDE, mirror reflectivity and total transmission (values obtained through measurements or from manufacturer specifications).
\item Mirror Layout - Missing mirror facets at the time of observations were removed from the model (18 out of 86). In addition the mirrors had begun to show signs of degradation due to their prolonged exposure to the environment at Adlershof. This will be evaluated when comparing the observed proton rate of the telescope (See Section \ref{s:bias}).
\item Point Spread Function - The PSF was measured on site using a white target in front of the focal plane and a CCD camera mounted on the telescope structure. Images of stars were taken and the 80\% containment angle was measured. The average value obtained was 0.218 deg, about three times worse than expected due to the non-smooth distribution of light in the image, most likely originating from misaligned mirrors (see middle panel of Figure \ref{fig:fig3}). The spread, but not the structure, was matched in the simulations.
\item Shadowing - In addition to the UV filter, a baffle was also mounted on the camera to reduce the amount of background light entering the camera, as can be seen in the right panel in Figure \ref{fig:fig3}. The effect of this on the shadowing on the camera was evaluated using the ray tracing software ROBAST \cite{robast}.
\end{itemize}

For the \texttt{CORSIKA} site simulation, the following parameters were adopted: Altitude of 37~m; MODTRAN atmospheric transmissivity model for tropical atmosphere; NRLMSISE-00  atmospheric density and refractive index model for CTA northern site; Magnetic field strength of H = 18.450~$\mu$T and Z = 45.351~$\mu$T, calculated from the Latitude and Longitude of the site location of $\phi =52.43^{\circ}$ N, $\lambda=13.54^{\circ}$ E and British Geological Survey World Magnetic Model.\footnote{\url{http://www.geomag.bgs.ac.uk/data_service/models_compass/wmm_calc.html}} While the atmospheric models are clearly not tuned to the site in Adlershof, they were the available models that extended to sea level. The simulated initiating particles were protons as it is expected that only background events were observed.

\subsection{Threshold Scan Comparison}
\label{s:bias}

\begin{figure}
\centering
\begin{tabular}{c}
 \includegraphics[width=0.5\textwidth]{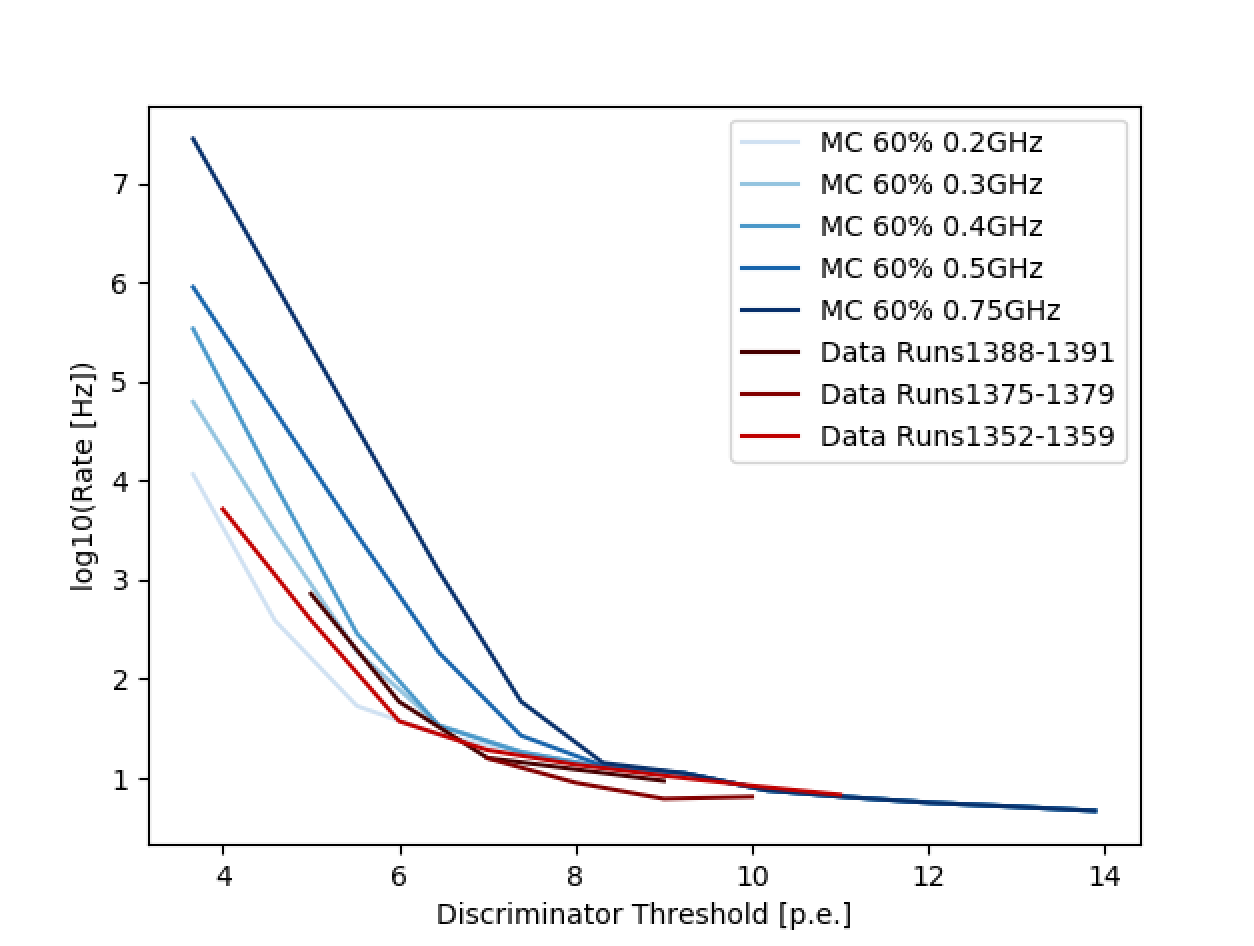} \\
\end{tabular}
\caption{Threshold scan for data in red and simulations in blue. For the MC simulation a range of NSB values is shown and only one mirror reflectivity value is shown (60\%). The data matches well to a NSB rate between 0.2 and 0.3 GHz (NSB rates provided in the legend). }
\label{fig:bais}
\end{figure}

During the on-sky tests at Adlershof, one of the frequent measurements was that of the threshold scan. This is a series of measurements of the trigger rate as a function of trigger threshold, similar to that performed in Section \ref{sec:trig}, which would be used to choose a safe observational trigger level. For this study, three on-sky runs that were carried out under similar observational conditions were used (pointing at dark sky spot, open shutter, nominal HV, internal trigger).

For the simulations, the trigger rate from proton showers and NSB have to be calculated separately. For this study $1.25\times10^{6}$ proton showers were simulated with \texttt{CORSIKA} between the energy range of 80~GeV and 50~TeV with an energy spectrum of $E^{-2}$. Showers were set to originate from a field of view of radius 10$^{\circ}$ centered 20$^{\circ}$ from zenith and were scattered at observation level in an area of 600~m radius. For the telescope simulation, the model used in Section \ref{sec:testbench} was adopted with the changes reported at the start of this Section. In addition, the NSB rate, the mirror reflectivity (degradation), and the trigger threshold were varied to find matching values to the data.

The proton rate is calculated by finding the trigger efficiency as a function of energy and using the following expression,

\begin{equation}
R_{\mathrm{proton}} = S \cdot \Omega \int_{E_{min}}^{E_{max}} \phi(E)  \cdot \frac{N(E)_{trig}}{N(E)_{sim}}~dE, 
\label{E:rp}
\end{equation}
where the simulated area is defined as $S = \pi r^2$ with r = 600~m and the simulated solid angle as $\Omega = 2 \pi \cdot (1 - \cos(\theta))$ with $\theta =10$ degrees. The proton flux, $ \phi(E)$, was taken from Ref. \cite{ctamc} and is defined as $\phi(E) = 9.6\times 10^{-2} \cdot \left(E/TeV\right)^{-2.70} \textrm{ TeV$^{-1}$~s$^{-1}$~m$^{-2}$ sr$^{-1}$}.$

For the NSB simulation, a \texttt{CORSIKA} file containing $10^5$ events with no Cherenkov light is used. The rate is calculated using ratio of triggered events to an equivalent simulated observation time, determined by the number of simulated events and the width of the readout window. Once the proton and NSB rates are determined, they are added to obtain the complete threshold scan. The results of this can be seen in Figure \ref{fig:bais} where it was found that a mirror degradation down to 60\% was required to match the proton spectrum. This might not all be due to the weathering of the mirrors, but could also encompass loss of efficiency in other parts of the system or the use of not ideal atmospheric models. Ideally for a more reliable result, images from muon rings are used to measure the total throughput efficiency, however not enough events were recorded for this analysis. For the NSB it was found that a value of 0.3~GHz was required to match the data. This value is in agreement with measurements that were taken earlier in the campaign.

\subsection{Hillas Parameter Comparison}
\label{s:hillas}

With updates to the telescope model obtained in the previous sections, a full simulation was performed using $25\times10^6$ proton showers between 80~GeV and 100~TeV. The on-sky data were calibrated using per pixel gain measurements recorded in the lab. Although interleaved pedestals were available for some of the data sets, for simplicity they were instead estimated from the average over the events for each run, using the first 10 samples in each waveform. The data were processed with \texttt{ctapipe} \cite{ctapipe}, where the charge was extracted and the images were cleaned using a two-level threshold, where only core pixels with at least 10 p.e. and any boundary pixels with at least 6 p.e. were kept, discarding any images that had less than 4 pixels remaining. 


\begin{figure}
\centering
\includegraphics[width=\textwidth]{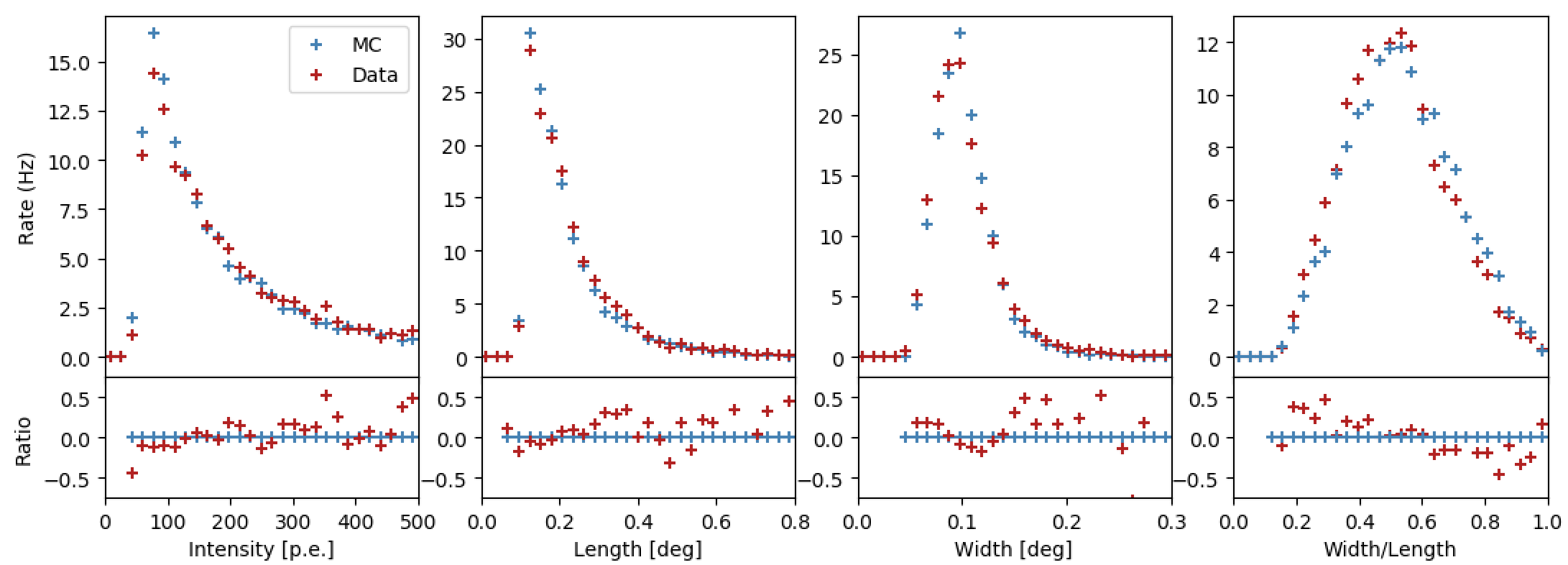} 
\caption{Comparison of selected Hillas parameters. The results from the simulation have been re-weighted to the assumed background proton spectrum from \cite{ctamc} and the rate per bin has been calculated for each set of results.}
\label{fig:hillas}
\end{figure}

For both MC and data, the cleaned images were fit with an ellipsoid in order to extract the Hillas parameters \cite{hillas} and the distributions were normalised to the expected rate. For the simulations, equation (\ref{E:rp}) was used to determine the rate. For the on-sky data the rate was calculated using the observation time of 26.65 min. In addition, simulated events are weighted before the construction of the histograms as $\frac{\phi_p}{\phi_{mc}}E^{\Gamma_{p}-\Gamma_{sim}}$ where $\Gamma_{p}$ is the assumed proton spectral index (-2.7),  $\Gamma_{sim}$ is the simulated spectral index (-2), and $\phi_{mc}$ is the simulated flux normalisation. The resulting distributions can be seen in Figure \ref{fig:hillas}, where a good agreement is found.

\section{Conclusion}

In this paper an overview of the main results produced during the model validation of NectarCAM has been shown. Concerning the comparison of results from the lab tests, both the readout and the trigger show good agreement. For the on-sky results, even though the conditions were not ideal for the observation of Cherenkov air-showers, an agreement was found through the scaling of the NSB (to 0.3~GHz) and the mirror degradation (down to 60\% of the original model). From these results, it is concluded that a good understanding of the camera has been achieved. 

\bigskip

\noindent
\textbf{Acknowledgements}: We gratefully acknowledge financial support from the agencies and organizations listed here:             
www.cta-observatory.org/consortium\_acknowledgments

%

%

\clearpage
\section*{Full Authors List: \Coll\ Collaboration}


\scriptsize
\noindent
H.~Abdalla\textsuperscript{1}, H.~Abe\textsuperscript{2},
S.~Abe\textsuperscript{2}, A.~Abusleme\textsuperscript{3},
F.~Acero\textsuperscript{4}, A.~Acharyya\textsuperscript{5}, V.~Acín
Portella\textsuperscript{6}, K.~Ackley\textsuperscript{7},
R.~Adam\textsuperscript{8}, C.~Adams\textsuperscript{9}
% ...
\\

\noindent
1 : Centre for Space Research, North-West University, Potchefstroom, 2520, South Africa
2 : Institute for Cosmic Ray Research, University of Tokyo, 5-1-5, Kashiwa-no-ha, Kashiwa, Chiba 277-8582, Japan
3 : Pontificia Universidad Católica de Chile, Av. Libertador Bernardo O'Higgins 340, Santiago, Chile
4 : AIM, CEA, CNRS, Université Paris-Saclay, Université Paris Diderot, Sorbonne Paris Cité, CEA Paris-Saclay, IRFU/DAp, Bat 709, Orme des Merisiers, 91191 Gif-sur-Yvette, France
5 : Centre for Advanced Instrumentation, Dept. of Physics, Durham University, South Road, Durham DH1 3LE, United Kingdom
6 : Port d'Informació Científica, Edifici D, Carrer de l'Albareda, 08193 Bellaterrra (Cerdanyola del Vallès), Spain
7 : School of Physics and Astronomy, Monash University, Melbourne, Victoria 3800, Australia
8 : Laboratoire Leprince-Ringuet, École Polytechnique (UMR 7638, CNRS/IN2P3, Institut Polytechnique de Paris), 91128 Palaiseau, France
9 : Department of Physics, Columbia University, 538 West 120th Street, New York, NY 10027, USA
% ...


\begin{thebibliography}{99}
\bibitem[Glicenstein et al.(2015)]{NectarCAM} Glicenstein, J.~F. et. al. 2015.\ NectarCAM : a camera for the medium size telescopes of the Cherenkov Telescope Array.\ 34th International Cosmic Ray Conference, 34.
\bibitem[Heck et al.(1998)]{corsika} Heck, D., Knapp, J., Capdevielle, J.~N., Schatz, G., Thouw, T.\ 1998.\ CORSIKA: a Monte Carlo code to simulate extensive air showers., Feb 1998, V + 90 p., TIB Hannover, D-30167 Hannover, Germany.
\bibitem[Bernl{\"o}hr(2008)]{simtel} Bernl{\"o}hr, K.\ 2008.\ Simulation of imaging atmospheric Cherenkov telescopes with CORSIKA and sim\_ telarray.\ Astroparticle Physics 30, 149–158. doi:10.1016/j.astropartphys.2008.07.009
\bibitem[Tavernier et al.(2019)]{icrc2019} Tavernier, T., Glicenstein, J.~F., Brun, F.\ 2019.\ Status and performance results from NectarCam - a camera for CTA medium sized telescopes.\ 36th International Cosmic Ray Conference, 36.
\bibitem{ctapipe} Kosack, K., Watson, J., Jacquemier, J., Nöthe, M., Mitchell, A., Neise, D, et. al., 2020, May 21, cta-observatory/ctapipe: Version v0.8.0. Zenodo. http://doi.org/10.5281/zenodo.3837306
\bibitem[Okumura et al.(2016)]{robast} Okumura, A., Noda, K., Rulten, C.\ 2016.\ ROBAST: Development of a ROOT-based ray-tracing library for cosmic-ray telescopes and its applications in the Cherenkov Telescope Array.\ Astroparticle Physics 76, 38–47. doi:10.1016/j.astropartphys.2015.12.003
\bibitem[Bernl{\"o}hr et al.(2013)]{ctamc} Bernl{\"o}hr, K. et. al. 2013.\ Monte Carlo design studies for the Cherenkov Telescope Array.\ Astroparticle Physics 43, 171–188. doi:10.1016/j.astropartphys.2012.10.002
\bibitem[Hillas(1985)]{hillas} Hillas, A.~M.\ 1985.\ Cerenkov Light Images of EAS Produced by Primary Gamma Rays and by Nuclei.\ 19th International Cosmic Ray Conference, Volume 3 3, 445.


\end{thebibliography}
\end{document}